\documentclass{PoS}

\usepackage{microtype,graphicx,amsmath,amssymb,wrapfig}

\let\OLDthebibliography\thebibliography
\renewcommand\thebibliography[1]{
  \OLDthebibliography{#1}
  \setlength{\parskip}{0pt}
  \setlength{\itemsep}{0pt plus 0.3ex}
}

\title{Generalized distribution amplitudes and
gravitational form factors for pion}

\ShortTitle{Generalized distribution amplitudes and
gravitational form factors for pion}

\author{\speaker{Qin-Tao Song}\\
 KEK Theory Center, Institute of Particle and Nuclear Studies, KEK,\\
           and Department of Particle and Nuclear Physics,
 Graduate University for Advanced Studies \\
 (SOKENDAI),       Ooho 1-1, Tsukuba, Ibaraki, 305-0801, Japan\\ 
 E-mail: \email{qintao@post.kek.jp}
 }

\abstract{Generalized parton distributions (GPDs) have been investigated in the deeply virtual Compton scattering (DVCS) to solve the proton spin puzzle. On the other hand, the generalized distribution amplitudes (GDAs) can be studied in the two-photon process $\gamma^* \gamma \rightarrow h \bar{h}$ which is accessible at KEKB. Namely, the GDAs are the $s$-$t$ crossed quantities of the GPDs. In 2016, the differential cross section of the process $\gamma^* \gamma \rightarrow \pi^0 \pi^0$ was measured by the Belle collaboration in the $e^+e^-$collision , so that the pion GDAs can be obtained by analyzing the Belle data. Expressing the GDAs with a few parameters, we determined the GDAs by a $\chi^2$ analysis . The form factors of the quark energy-momentum tensor are obtained from the determined GDAs. Then, we calculated the mass radius as 0.56-0.69 fm for the pion by using the form factor. This is the first study on gravitational form factors and radius of hadrons from actual experimental measurements . The Belle II will start taking data in 2018 by the upgraded SuperKEKB. Therefore, much accurate data are expected for the pion, and other hadron-pair productions will be measured in the near future. The GDA studies are valuable for understanding not only the 3D structure but also gravitational properties of hadrons.}

\FullConference{XXVI International Workshop on Deep-Inelastic Scattering and Related Subjects (DIS2018)\\
		16-20 April 2018\\
		Kobe, Japan}

\begin{document}

\section{Introduction}
The proton spin puzzle was observed by the European Muon Collaboration in 1980s, and it indicates that only a small fraction of the proton spin is carried by quarks in the proton. In order to study the proton spin puzzle, generalized parton distributions (GPDs) are widely investigated. In the deeply virtual Compton scattering (DVCS), the soft part is expressed by the GPDs. One can also study GPDs by other processes, such as  deeply virtual meson production (DVMS) 
process \cite{Favart:2015umi}, the exclusive hadoronic $2 \to 3$ reaction
\cite{Kumano:2009he}, and exclusive pion-induced Drell-Yan process \cite{Sawada:2016mao}. Another type of three-dimensional structure functions are generalized distribution amplitudes (GDAs), and they are the $s$-$t$ crossed quantities of GPDs. Therefore, we could obtain the GPD information by studying the GDAs, and they can be investigated by the two-photon process $\gamma^* \gamma \rightarrow h \bar{h}$ \cite{Diehl:1998dk,Diehl:2000uv, Polyakov:1998ze}. In the two photon process, the GDAs describe the amplitude of $q \bar{q} \rightarrow h \bar{h}$. 
The GPDs and GDAs carry important information of hadrons, such as form factors, 
parton distribution functions , distribution amplitudes and angular momentum of partons.

The GPDs are now studied at the Thomas Jefferson National Accelerator Facility (JLab) and at the European Organization for Nuclear Research (CERN) by $ep \rightarrow e \gamma p $ process, 
and they can be also investigated at at Japan Proton Accelerator Research Complex (J-PARC) 
through the exclusive pion-induced Drell-Yan process in future. 
As for the GDAs, the two-photon process $\gamma^* \gamma \rightarrow h \bar{h}$ is accessible at KEKB.
In this work, the GDA of pion is obtained by analyzing the Belle data $\gamma \gamma^* \rightarrow \pi^0 \pi^0$ \cite{Masuda:2015yoh}. 
In the near future, the Belle II will start to collect data with the higher luminosity super KEKB, 
and our GDA study could be used to explore the GDAs of other hadrons with more precise measurements of $\gamma \gamma^* \rightarrow h \bar{h}$.

\section{GDAs in the two-photon process}

\begin{figure}[htb]
\vspace{-0.5em}
\centering
\includegraphics[width=0.5\textwidth]{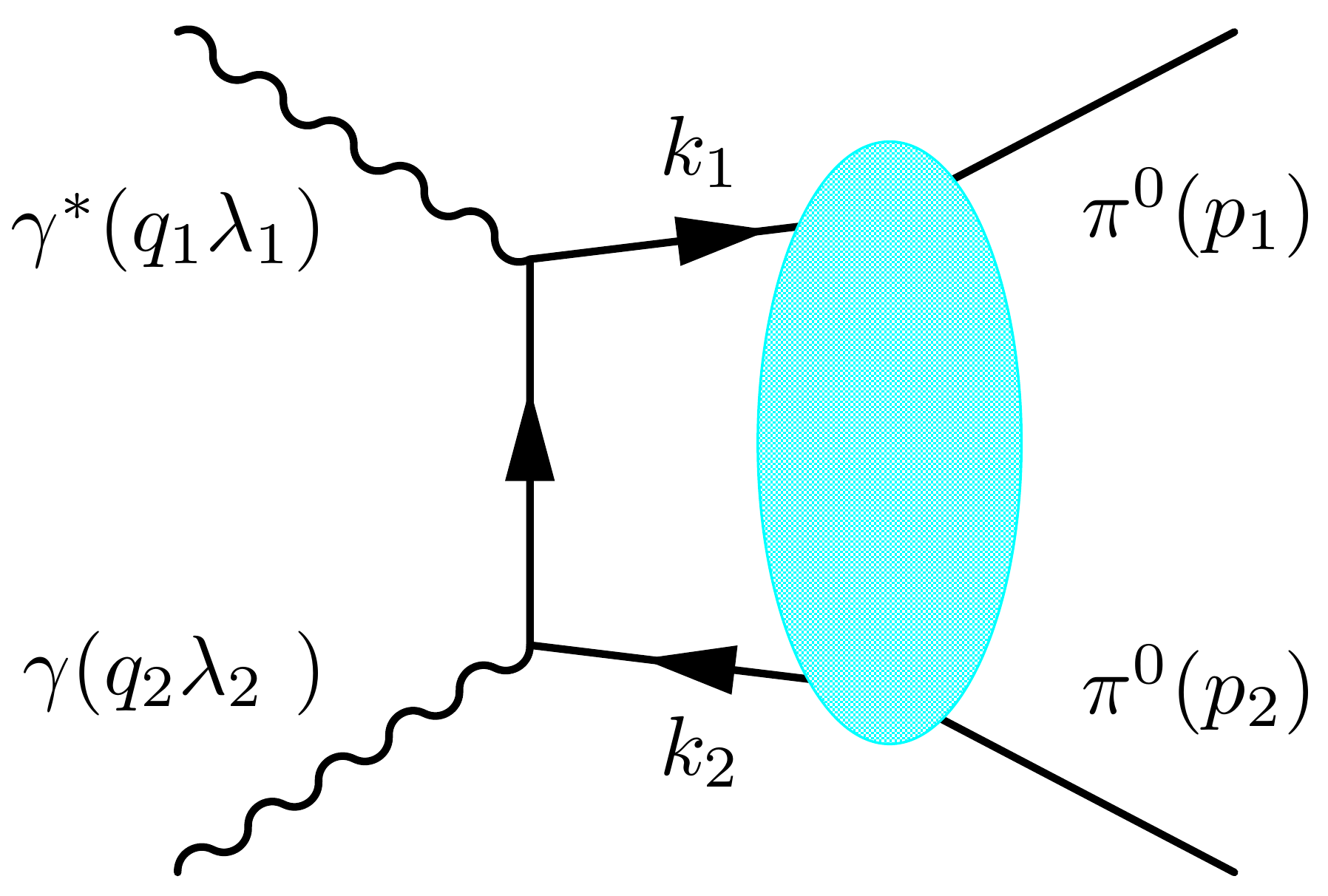}
\vspace{0.0em}
\caption{The two photon process $\gamma^* \gamma \rightarrow \pi^0 \pi^0$, and the soft part is the GDA of pion.}
\label{fig:pion-gda}

\end{figure}

In the process $\gamma^* \gamma \rightarrow \pi^0 \pi^0$ shown in Fig.\,\ref{fig:pion-gda}, if $Q^2=-q^2_1$ of the virtual photon is large enough, the amplitude can be fractorized into a soft part and a hard part. The hard part is the process $\gamma^* \gamma \rightarrow q \bar{q}$, with produced collinear and
on-shell quark, and a soft part is the production of the hadron pair
$h \bar{h}$ from a $q \bar{q}$. This soft part is called GDA \cite{Diehl:1998dk,Diehl:2000uv, Polyakov:1998ze}, and it is expressed as
\begin{align} 
& \Phi_q^{\pi^0 \pi^0} (z,\zeta,W^2) 
= \int \frac{d y^-}{2\pi}\, e^{i (2z-1)\, P^+ y^- /2}
  \langle \, \pi^0 (p) \, \pi^0 (p') \, | \, 
 \overline{q}(-y/2) \gamma^+ q(y/2) 
  \, | \, 0 \, \rangle \Big |_{y^+=\vec y_\perp =0} \, ,
\label{eqn:gda-def}
\end{align}
where $W^2=(p_1+p_2)^2=P^2$, $z=k_1^+/P^+$ and $z=p_1^+/P^+$ are defined. In Fig.\,\ref{fig:pion-gda},
there are three independent helicity amplitudes $A_{\lambda_1 \lambda_2}$, and the leading twist and leading order contribution is the one with same helicity for the incoming photons ($\lambda_1 =\lambda_2$). In the large $Q^2$ region, the higher-twist and higher-order amplitudes can be neglected. In this case, we can express the differential cross section of  $\gamma^* \gamma \rightarrow \pi^0 \pi^0$ with the GDAs \cite{Diehl:1998dk,Diehl:2000uv} defined in Eq. (\ref{eqn:gda-def}) as
\begin{align}
&d\sigma=\frac{1}{4}  \alpha^2  \pi \frac{\sqrt{1-\frac{4m^2}{s}} }{ Q^2+s}  |A_{++}|^2  \sin\theta  d\theta, %\\
&A_{++}=\sum_q \frac{e_q^2}{2} \int^1_0 dz \frac{2z-1}{z(1-z)} \Phi^{\pi^0 \pi^0}_q(z, \xi, W^2).  %\nonumber .
\label{eqn:amp2}
\end{align}

 In order to analyze the Belle data \cite{Masuda:2015yoh}, one needs to know the general GDA expression for pion.
In very large $Q^2$ limit, we have the asymptotic form of the GDA \cite{Diehl:1998dk,Diehl:2000uv}
\begin{align}
&\sum_q \Phi^{\pi^0\pi^0}_q(z, \xi, W^2)=18n_fz(1-z)(2z-1)[\tilde{B}_{10}(W)+\tilde{B}_{12}(W)P_2(cos\theta)],  \nonumber \\
&\tilde{B}_{nl}(W)=\bar{B}_{nl}(W)exp(i\delta_l), \zeta  = \frac{1+\beta \cos\theta}{2}, \beta=\sqrt{1-\frac{4m^2}{s}},
\label{eqn:gda}
\end{align}
where $\tilde{B}_{10}(W)$ and  $\tilde{B}_{12}(W)$ indicate the S-wave and D-wave productions of the pion pair, respectively.
In Eq. (\ref{eqn:gda}),  $\delta_0$ and $\delta_2$  are the $\pi\pi$ elastic scattering phase shifts \cite{Bydzovsky:2014cda, Nazari:2016wio} in
the isospin=0 channel below the $KK$ threshold.  Above the threshold, the additional
phase is introduced for S-wave  phase shift in this GDA analysis.

In the two-photon process, $\pi^0 \pi^0$ can be produced through intermediate meson
state $h$ \cite{Anikin:2004ja}, $\gamma^* \gamma \rightarrow h  \rightarrow \pi^0 \pi^0$. The intermediate resonances can be $f_0(500)$ for S wave and $f_0(1270)$ for D wave, and those resonance contributions play an important role in the resonance region.

\section{GDAs analysis of the Belle data}
\begin{figure}[hbt]

\vspace{0.2em}
\centering
\includegraphics[width=0.5\textwidth]{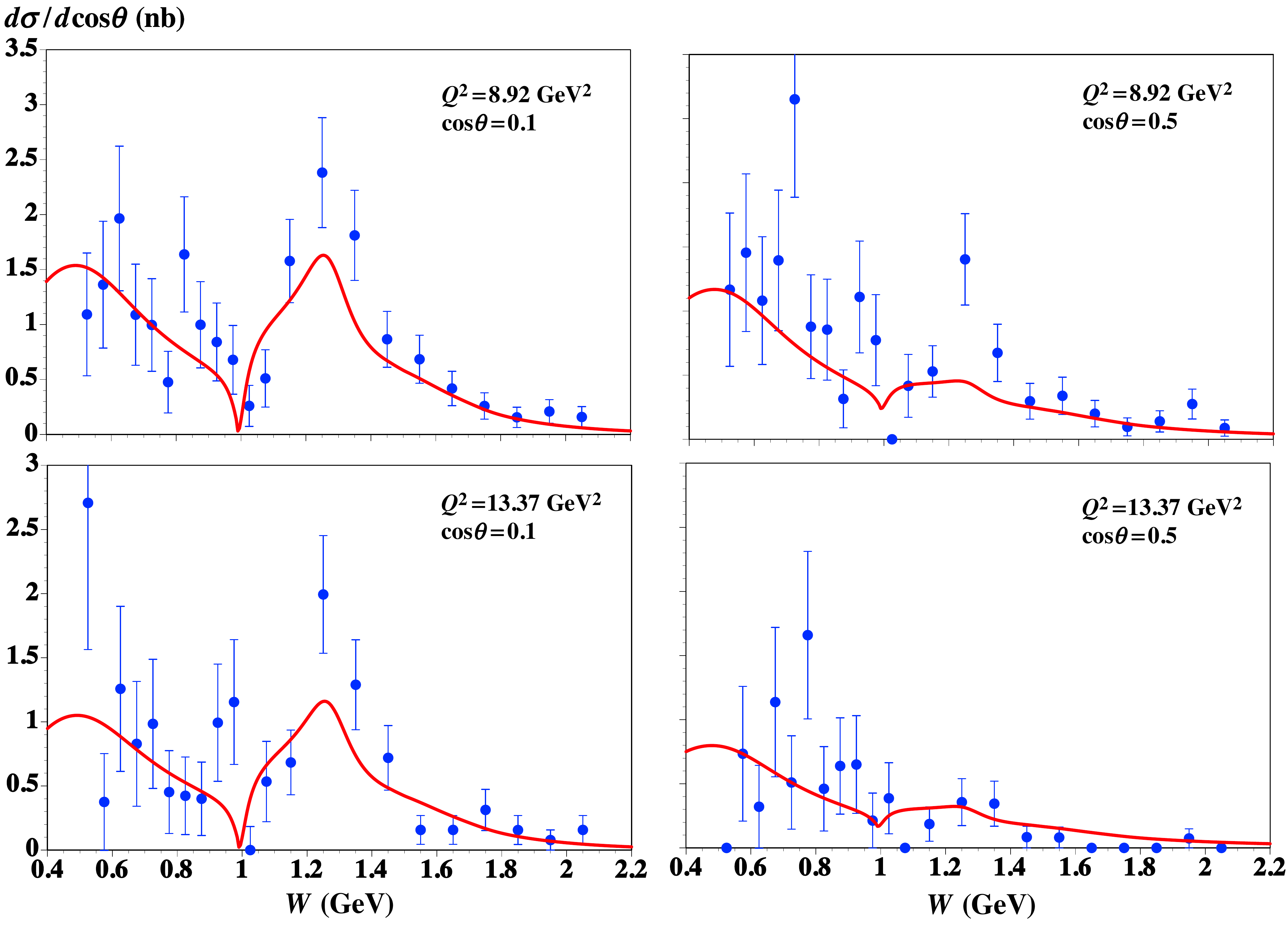}
\vspace{-0.5em}
\caption{The $W$ dependence of the differential cross section of  $\gamma^* \gamma \rightarrow \pi^0 \pi^0$ (in units of nb) in comparison with Belle data. The $Q^2$ values are fixed as 8.92 GeV and 13.37 GeV, and $\cos\theta$ is fixed as 0.1 and 0.5 \cite{Kumano:2017lhr}.}
\label{fig:cross}
\end{figure}

We adopt a simple expression of GDA to analyze Belle data \cite{Kumano:2017lhr, Kawamura:2013wfa}:
\begin{align}
&\Phi_q^{\pi^0 \pi^0}(z, \xi, W^2)=N_h z^\alpha(1-z)^\alpha(2z-1)  [\tilde{B}_{10}(W)+\tilde{B}_{12}(W)P_2(cos\theta)],  \nonumber \\
%&[B_{10}(W)+B_{12}(W)P_2(2\zeta-1)]= [\tilde{B}_{10}(W)+\tilde{B}_{12}(W)P_2(cos\theta)] ,  \nonumber \\ 
%&\tilde{B}_{10}(W)=B_{10}(W)-\frac{1-\beta^2}{2}B_{12}(W),   \nonumber \\ 
%&\tilde{B}_{12}(W)=\beta^2B_{12}(W),P_2(2\zeta-1)=1-6\zeta( \zeta-1), \nonumber \\ 
& \tilde{B}_{10}(W)=\left \{ \frac{-3+\beta^2}{2}\frac{10R_{\pi}}{9n_f} F_h(W^2) +     \frac{5g_{f_0 \pi\pi} f_{f_0} }{3 \sqrt{2} \sqrt{[(M^2_{f_0}-W^2)^2+\Gamma^2_{f_0} M^2_{f_0} ]}}     \right \} e^{i\delta_0}     , \nonumber \\ 
&  \tilde{B}_{12}(W)=\left \{ \beta^2 \frac{10R_{\pi}}{9n_f} F_h(W^2) + \beta^2 \frac{10g_{f_2\pi\pi} f_{f_2} M^2_{f_2} }{9 \sqrt{2} \sqrt{ (M^2_{f_2}-W^2)^2+\Gamma^2_{f_2} M^2_{f_2} }}  \right \}   e^{i\delta_2}.
\label{eqn:gdacb}
\end{align}
Here, resonance effects of $f_0(500)$ and $f_2(1270)$ are introduced,  and $f_0(980)$ is not included since it is not seen in the  the differential cross section of $\gamma^* \gamma \rightarrow \pi^0 \pi^0$.
 In the Eq. (\ref{eqn:gdacb}), 5 parameters are introduced to analyze the Belle data.
  $\delta_0$ and $\delta_2$ are the $\pi\pi$ scattering phase shifts \cite{Bydzovsky:2014cda, Nazari:2016wio}.
In the asymptotic limit,  the parameter $\alpha$ is predicted as $\alpha=1$. $R_{\pi}$ is the momentum fraction carried by quarks for the pion. 
The function $N_h$ is dependent on $\alpha$, and it ensures that the following sum rule is satisfied \cite{Polyakov:1998ze}
\begin{align}
\int_0^1 dz (2z -1) \, 
\Phi_q^{\pi^0 \pi^0} (z,\,\zeta,\, 0) =-2R_{\pi}\zeta(1-\zeta ).
\label{eqn:integral-over-z}
\end{align}
We explain the details of our analysis results by using Eq. (\ref{eqn:gdacb}) \cite{Kumano:2017lhr}.
 The values of parameters are obtained by analyzing the Belle data. We have $\chi^2/\text{d.o.f.}=1.09$ in this analysis, which gives  a resonable description of the Belle data. In Fig. \ref{fig:cross}, the differential cross section of  $\gamma^* \gamma \rightarrow \pi^0 \pi^0$ is shown with fixed $Q^2$ and $\cos \theta $ in comparison with Belle data, and the resonance effect of $f_2(1270)$  is clearly seen around $W=1.2$ GeV.

\begin{figure}[htb]
\vspace{0.2em}
\centering
\includegraphics[width=0.5\textwidth]{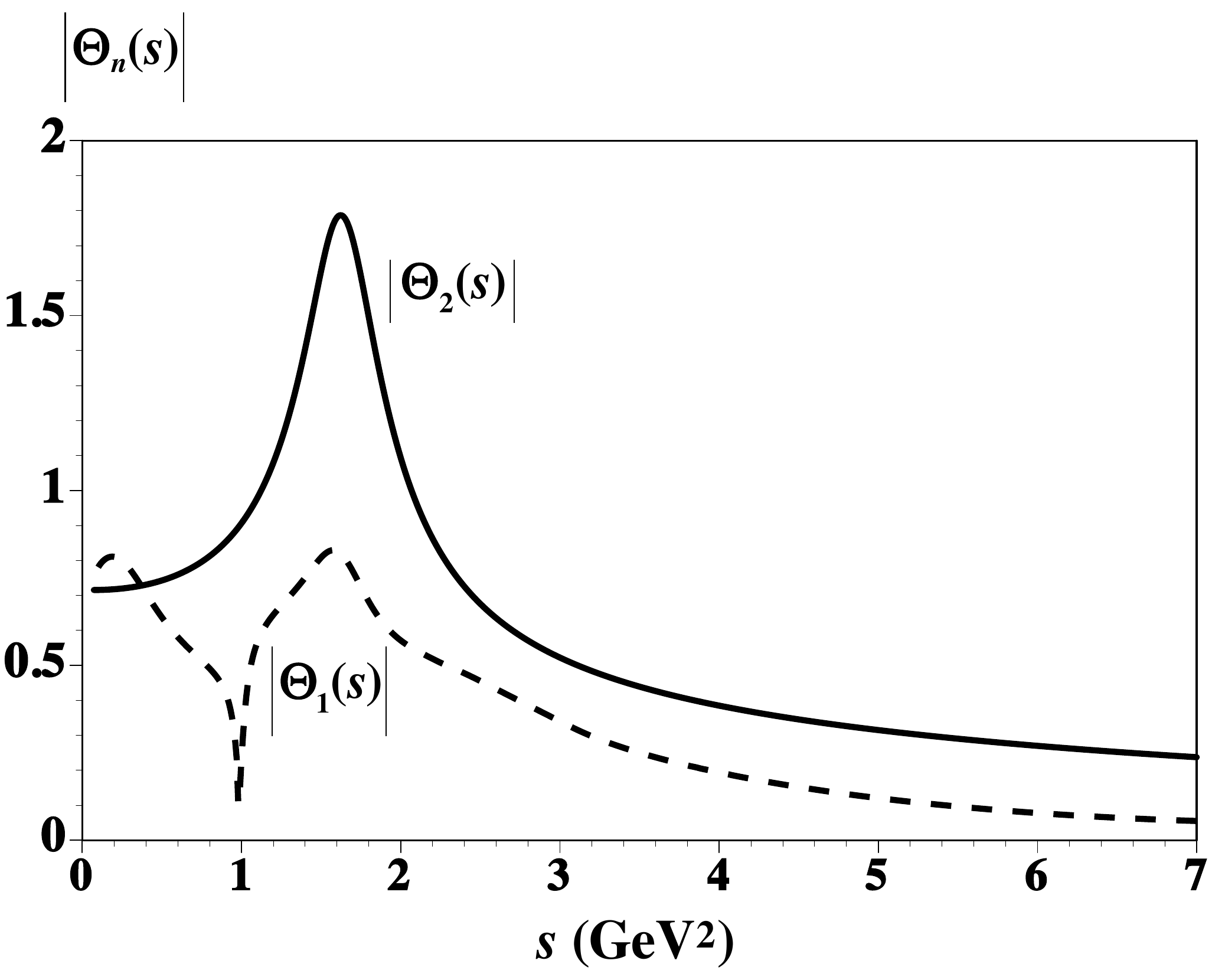}
\vspace{-0.5em}
\caption{The gravitational form factors $\Theta_{1}(s)$ and  $\Theta_{2}(s)$  in the timelike region for pion \cite{Kumano:2017lhr}.  }
\label{fig:theta}
\end{figure}

With the obtained GDA of pion, we can also study the energy-momentum form factors in the timelike region. The energy-momentum tensor of pion  $T_q^{\mu \nu}$ can be connected to the GDA by the following equation \cite{Polyakov:1998ze}
\begin{align}
& \int_0^1 dz (2z -1) \, 
\Phi_q^{\pi^0 \pi^0} (z,\,\zeta,\,W^2) = \frac{2}{(P^+)^2} \langle \, \pi^0 (p) \, \pi^0 (p') \, | \, T_q^{++} (0) \,
       | \, 0 \, \rangle .
\label{eqn:integral-over-z}
\end{align}
The definition of energy-momentum tensor reads
\begin{align}
& \langle \, \pi^0 (p) \, \pi^0 (p') \, | \sum_q  \, T_q^{\mu\nu} (0) \, | \, 0 \, \rangle 
= \frac{1}{2} 
  \left [ \, \left ( s \, g^{\mu\nu} -P^\mu P^\nu \right ) \, \Theta_{1} (s)
                + \Delta^\mu \Delta^\nu \,  \Theta_{2} (s) \,
  \right ],
\label{eqn:emt-ffs-timelike-0}
\end{align}
where  $\Theta_{1}$ and  $\Theta_{2}$ are the gravitational form factors. $\Theta_{1}$ is related to the mass or energy, and $\Theta_{2}$ is related to mechanic (pressure and sheer force).
From Eqs.\,(\ref{eqn:integral-over-z}) and (\ref{eqn:emt-ffs-timelike-0}) ,
 the gravitational form factors for the pion can be calculated as
 \begin{align}
\Theta_{1} (s) 
= \frac{3}{5} ( \widetilde B_{12} (W^2)-2\widetilde B_{10} (W^2)  ),
\Theta_{2} (s)  
= \frac{9}{5 \, \beta^2} \widetilde B_{12} (W^2) .
\label{eqn:emt-ffs-gdas}
\end{align}
The timelike form factors $\Theta_{1}(s)$ and  $\Theta_{2}(s)$ are shown in Fig. \ref {fig:theta}. $\Theta_{1}(s)$ contains both S wave and D wave, and the S wave interferes with the D wave .
However, $\Theta_{2}(s)$ is purely D wave.

\begin{figure}[hbt]
\centering
\includegraphics[width=0.49\textwidth]{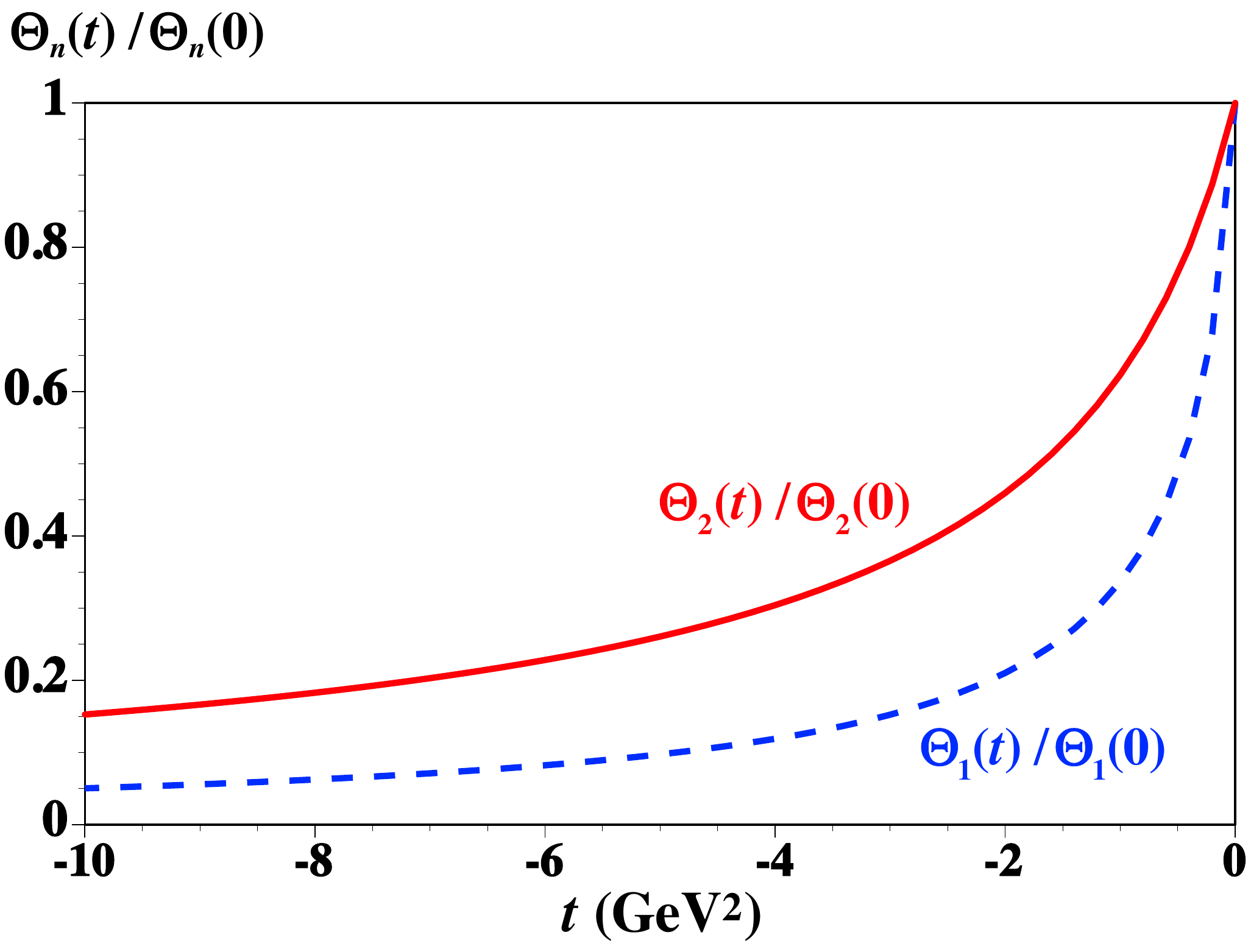}
\includegraphics[width=0.49\textwidth]{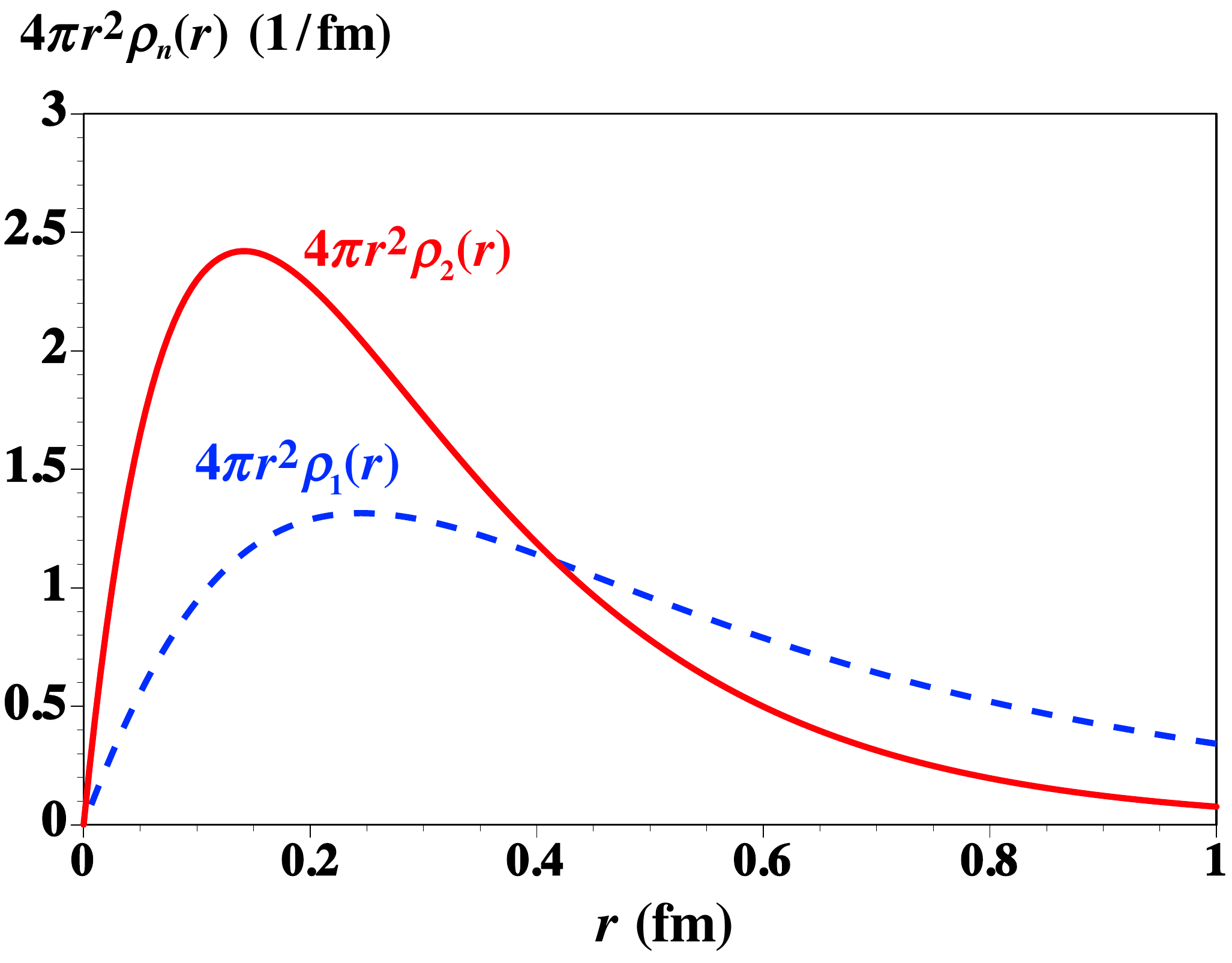}
\caption{Left panel: the gravitational form factors $\Theta_{1}(t)$ and  $\Theta_{2}(t)$ in the spacelike region for pion. Right panel: $\rho_1(r)$ and $\rho_2(r)$ are the Fourier transforms of the spacelike gravitational form factors $\Theta_{1}(t)$ and  $\Theta_{2}(t)$, respectively \cite{Kumano:2017lhr}.}
\label{fig:rho}
\end{figure}

In order to study the spacelike gravitational form factors, the dispersion relation is needed \cite{form-factor-dispersion}:
\begin{align}
F^h (t) & = \int_{4 m_h^2}^\infty \frac{ds}{\pi} 
            \frac{{\rm Im}\, F^h (s)}{s-t-i\varepsilon}.
\label{eqn:dispersion-form-1}
\end{align}
The spacelike gravitational form factors $\Theta_{1}(t)$ and  $\Theta_{2}(t)$ are shown in left panel of Fig. \ref{fig:rho}, and they are normalized as 1 at $t=0$. 
$\Theta_{1}(t)$ decreases more rapidly than $\Theta_{2}(t)$ as $|t|$ increases.
We also make the plot for $\rho_1(r)$ and $\rho_2(r)$ in right panel of Fig. \ref{fig:rho}, and they are just the Fourier transforms of the factors $\Theta_{1}(t)$ and  $\Theta_{2}(t)$. The distribution $\rho_1(r)$  is concentrated at the region $r=0.1-0.3$ fm, however, the distribution $\rho_2(r)$ is much flatter.

The three-dimensional rms radii are also calculated by taking the slopes of  gravitational form factors $\Theta_{1}(t)$ and  $\Theta_{2}(t)$ at $t=0$:
\begin{align}
\langle \,  r^2 \, \rangle _h
=  \left. 6 \, \frac{F ^h (t)}{dt} \right|_{|t| \to 0} 
= 6 \int_{4 m_\pi^2}^\infty ds \frac{{\rm Im} \, F^h(s)}{s^2}.
\label{eqn:3D-radius-2}
\end{align}
$\Theta_{1}$ is reveals information on the mass distribution, and we obtain mass radius $\sqrt {\langle r^2 \rangle _{\text{mass}}}=0.69$ fm.   $\Theta_{2}$ is related to mechanic distribution (pressure and sheer force), so the mechanic radius is $\sqrt {\langle r^2 \rangle _{\text{mech}}}=1.45$ fm \cite{Kumano:2017lhr}.
In our analysis we introduce the additional phase for S-wave above the $KK$
threshold. However, the additional phase could be added to D-wave phase above
the threshold, in this case, we have $\sqrt {\langle r^2 \rangle _{\text{mass}}}=0.56$ fm and $\sqrt {\langle r^2 \rangle _{\text{mech}}}=1.56$ fm. Therefore, it is better to express the gravitational radii as \cite{Kumano:2017lhr}
\begin{align}
\sqrt {\langle r^2 \rangle _{\text{mass}}} 
=  0.56 \sim 0.69 \, \text{fm}, \, 
\sqrt {\langle r^2 \rangle _{\text{mech}}} 
 = 1.45 \sim 1.56 \, \text{fm} .
\label{eqn:g-radii-pion-range}
\end{align}
The mechanical radius is much larger than the mass radius, and the mass radius is slightly smaller than or similar with the charge radius of pion $\sqrt {\langle r^2 \rangle _{\text{charge}}}=0.672 \pm 0.008$ fm \cite{Patrignani:2016xqp}. There are other recent works on the gravitational form factors \cite{Burkert:2018bqq}.

\section{Summary}
The GDAs are the $s$-$t$ crossed quantities of GPDs, and GDAs can provide us another way to study GPDs.
Both GPDs and GDAs carry important information of the hadrons. In 2016, the Belle collaboration measured the differential cross section of the process $\gamma^* \gamma \rightarrow \pi^0 \pi^0$. In this work, we analyzed the Belle data by using a simple GDA expression for pion, and the obtained GDA can describe experimental data very well.
Moreover, the GDAs are also used to study energy-momentum tensor of hadron.
The timelike energy-momentum form factors of pion are calculated from the GDA of pion, and the spacelike energy-momentum form factors $\Theta_{1}(t)$ and  $\Theta_{2}(t)$  are obtained by the dispersion relation.
This is the first finding on gravitational radii of hadrons from actual
experimental measurements: we obtained the mass radius (0.56-0.69fm) and
the mechanical radius (1.45-1.56fm). In the near future, the GDAs of other hadrons can also be studied with more precise data from the forthcoming Belle II.

\acknowledgments
\noindent Q.-T. S is supported by the MEXT Scholarship for foreign students 
through the Graduate University for Advanced Studies.

\bibliographystyle{JHEP}
\bibliography{SongDIS2018}

\end{document}